\documentclass{emulateapj}

\newcommand{\etal}{{\it et\thinspace al.}\ }
\newcommand{\kms}{km\thinspace s$^{-1}$}
\newcommand{\simlt}{\ {\raise-.5ex\hbox{$\buildrel<\over\sim$}}\ }

\submitted{Submitted 20 June 2005; Accepted 29 July 2005 -- To appear in the 
2005 December {\it Astronomical Journal}}

\lefthead{Jangren \etal}
\righthead{KPNO International Spectroscopic Survey. V.}

\begin{document}

\title{The KPNO International Spectroscopic Survey.  \\ V. H$\alpha$-selected Survey List 3.}

\author{Anna Jangren, John J. Salzer, Vicki L. Sarajedini\altaffilmark{1}, Caryl Gronwall\altaffilmark{2}, Jessica K. Werk\altaffilmark{3}, Laura Chomiuk\altaffilmark{4} }
\affil{Astronomy Department, Wesleyan University, Middletown, CT 06459; 
anna@astro.wesleyan.edu, slaz@astro.wesleyan.edu}

\author{J. Ward Moody}
\affil{Department of Physics \& Astronomy, Brigham Young University, Provo, UT 84602; jmoody@astro.byu.edu}

\and

\author{Todd A. Boroson}
\affil{National Optical Astronomy Obs., P.O. Box 26732, Tucson, AZ 85726; tyb@noao.edu}

\altaffiltext{1}{Current address: Astronomy Department, University of Florida,
Gainesville, FL, 32611; vicki@astro.ufl.edu.}

\altaffiltext{2}{Current address: Department of Astronomy \& Astrophysics, 
Pennsylvania State University, University Park, PA 16802; caryl@astro.psu.edu.}

\altaffiltext{3}{Current address: Department of Astronomy,
University of  Michigan, Ann Arbor, MI 48109; jwerk@umich.edu}  

\altaffiltext{4}{Current address: Department of Astronomy, University of Wisconsin at 
Madison, 475 N. Charter St, Madison, WI 53706; chomiuk@astro.wisc.edu}

\begin{abstract}
The KPNO International Spectroscopic Survey (KISS) is an objective-prism
survey designed to detect extragalactic emission-line objects.  It combines many of the 
features of previous slitless spectroscopic surveys with the advantages of 
modern CCD detectors, and is the first purely digital objective-prism survey
for emission-line galaxies (ELGs).  Here we present the third list of ELG
candidates selected from our red spectral data, which cover the wavelength
range 6400 to 7200 \AA.  In most cases, the detected emission line is H$\alpha$.  
The current survey list covers the region of the NOAO Deep Wide-Field Survey (NDWFS).
This survey covers two fields; the first is $3 \times 3$ degrees square and located at 
RA = 14$^h$~30$^m$, $\delta$ = 34$\arcdeg$~30$\arcmin$ (B1950), the second is
$2.3 \times 4.0$ degrees and centered at RA = 2$^h$~7.5$^m$, $\delta$ = 
-4$\arcdeg$~44$\arcmin$.   A total area of 19.65 deg$^2$ is covered by the KISS data.  
A total of 261 candidate emission-line objects have been selected for inclusion 
in the survey list (13.3 per deg$^2$).  
We tabulate accurate coordinates and photometry for each source, as well as 
estimates of the redshift, emission-line flux and line equivalent width based on 
measurements of the digital objective-prism spectra.  The properties of the KISS 
ELGs are examined using the available observational data.  When combined with 
the wealth of multi-wavelength data already available for the NDWFS fields, the 
current list of KISS ELGs should provide a valuable tool for studying star-formation
and nuclear activity in galaxies in the local universe.
\end{abstract}

\keywords{galaxies: emission lines --- galaxies: Seyfert --- galaxies: starburst --- surveys}


\section{Introduction}

Surveys for galaxies containing active galactic nuclei (AGNs) or strong star-formation
activity have been an important area of extragalactic astronomy for decades.  Many fruitful
surveys have been carried out with wide-field Schmidt telescopes used in conjunction
with objective prisms.  An overview of previous surveys is given in Salzer \etal (2000),
along with a sampling of the types of applications that such surveys have for the study
of the extragalactic universe.

We have been carrying out a modern objective-prism survey for the past several years.
Called the KPNO International Spectroscopic Survey (KISS), it combines many of the
advantages of older surveys with the use of state-of-the-art CCD detectors, providing
superior depth and data quality.  The digital nature of KISS has many advantages over
the older photographic surveys of this type (e.g., Markarian 1967; Smith, Aguirre \& Zemelman
1976; MacAlpine, Smith \& Lewis 1977; Pesch \& Sanduleak 1983; Wasilewski 1983;
Markarian, Lipovetskii, \& Stepanian 1983; Zamorano \etal 1994; Popescu \etal 1996; 
Surace \& Comte 1998; Ugryumov \etal 1999).    Besides the obvious factors of higher
sensitivity and speed, we stress the importance of being able to measure the
completeness limits and selection function of the survey directly from the data used to
derive the catalogs of KISS emission-line galaxies (ELGs).  This is not possible with
photographic survey material, and makes KISS particularly useful for statistical studies
of galaxian activity in the nearby universe. 

The current survey lists  cover the area of the sky included in the NOAO Deep Wide-Field
Survey (NDWFS; Jannuzi \& Dey 1999, Jannuzi \etal in preparation, Dey \etal in preparation). 
NDWFS is a deep optical and NIR imaging survey carried out in two well separated fields.  
All optical data were taken on the NOAO 4-m telescopes in the BRI bandpasses, while JHK 
imaging was carried out on the KPNO 2.1-m telescope.  The fields were both covered to
a uniform depth of B $\approx$ 26.6 (and correspondingly deep in the other five bands).
We chose to observe these fields 
as part of KISS because of the expectation that they would become well observed
at many wavelengths as various groups studied the properties of the NDWFS galaxies.
While the primary science goals of the NDWFS focus on galaxies at redshifts well
beyond the filter-imposed redshift limit of KISS (z $\le$ 0.095), the volume covered 
by our survey is sufficiently large to provide a good-sized sample of star-forming galaxies 
and AGNs.   Bolstered by the large amount of data becoming available for the galaxies
in the NDWFS area at radio, FIR, NIR, UV, X-ray, and optical wavelengths, the KISS
ELGs should allow for a number of detailed statistical studies of activity in galaxies in 
the local universe.  While the KISS data are completely independent of the NDWFS
data, they can be used to complement and extend the usefulness of the latter.

This is the fifth paper in the KISS series.   The first presents a complete description of the
survey method, including a discussion of the survey data and its associated uncertainties  
(Salzer \etal 2000; hereafter Paper I).  The first and second survey lists of H$\alpha$-selected 
ELGs, informally referred to as the red survey, are given in Salzer \etal (2001; hereafter KR1)
and Gronwall \etal (2004; hereafter KR2), while the first list of [\ion{O}{3}]-selected galaxies 
(the blue survey) is found in Salzer \etal (2002; hereafter KB1).   
The current paper follows a format similar to KR1 and KR2; for the sake of brevity,
the reader is referred to KR1, KR2, and Paper I for many details.  The  observational data 
and image processing are described in Section 2, while the new list of ELG candidates
is presented in Section 3.  The properties of the new list of H$\alpha$-selected 
ELGs are described in Section 4, while our results are summarized in Section 5.


\section{Observations \& Reductions}

All survey data were acquired using the 0.61-meter Burrell Schmidt 
telescope\footnote{Observations made with the Burrell Schmidt telescope of 
the Warner and Swasey Observatory, Case Western Reserve University.}. 
The detector used for all data reported here was a 2048 $\times$ 4096 
pixel SITe CCD.  The CCD is identical to the one used for the KR2 list, however 
this is not the same CCD that was used for KR1 or KB1, giving a different image 
scale and field-of-view.  The CCD has 15-$\micron$ pixels, yielding an image 
scale of 1.43 arcsec/pixel at the Newtonian focus of the telescope. The overall 
field-of-view was 50 $\times$ 100 arcmin, and each image covered 1.37 square 
degrees.  The long dimension of the CCD was oriented north-south during our survey 
observations.  The red survey spectral data were obtained with a 4$\arcdeg$ 
prism, which provided a reciprocal dispersion of 17 \AA/pixel at H$\alpha$.   The 
spectral data were obtained through a special filter designed for the survey, which 
covered the spectral range 6400 -- 7200 \AA\ (see Figure 1 of Paper I for the filter 
transmission curve).

The two NDWFS fields each cover an area of 9 sq. deg.  The spring (Bo\"otes) field
is centered at RA = 14$^h$~30$^m$, Dec = 34$\arcdeg$~30$\arcmin$ (B1950).  It
consists of a 3 $\times$ 3 degree square field.  The fall (Cetus) field is a 2.3 $\times$ 4.0
degree area centered at RA = 2$^h$~7.5$^m$, Dec = -4$\arcdeg$~44$\arcmin$ (B1950).
The layout of the Bo\"otes field allowed us to cover the NDWFS area with two rows of
KISS fields, with four fields per row.  There is essentially zero overlap in declination
between the two rows of fields.  In addition, due to larger than normal pointing offsets
between the direct and spectral fields (see below), there are modest gaps between
some of the fields within a given row.  The net result is that the KISS data for the Bo\"otes
field only cover 8.08 sq. deg., rather than the full 9 sq. deg.  For the Cetus fields, we
again utilized two rows of KISS fields.  However, in this case there is substantial
overlap between the upper and lower rows, due to the fact that the declination extent
of this NDWFS region is smaller.  Furthermore, we needed 6 KISS fields per row to
cover the full 4.0 degrees of RA.  Despite the declination overlap, the Cetus KISS fields
cover a total area of 12.57 sq. deg., substantially larger than the area of the NDWFS
fields.  The total area covered by the KISS observations is 19.65 sq. deg.

As with our previous survey strips, we obtained images of each survey field both
with and without the objective prism on the telescope.  The images taken without
the prism (referred to as direct images) were obtained through standard B and V
filters.  The direct images were photometrically calibrated, and provide accurate 
astrometry and photometry for all sources in the survey fields.  We used uniform 
exposure times for all survey fields: 4 $\times$ 720 s for the objective-prism (spectral)
data, and 2 $\times$ 300 s for V and 1 $\times$ 600 s in B for the direct images.
The telescope was dithered by a small amount ($\sim$10 arcsec) between exposures.

Table \ref{table:tab1} lists the observing runs during which the current set of survey
fields were observed. The first column gives the UT dates of the run, while the second 
column indicates the number of nights on which observations were obtained.  At least 
some data were obtained on 12 of 14 scheduled nights (86\%).  The last two columns
indicate the number of direct and spectral images, respectively, obtained during each run. 
It was common practice to observe in both direct and spectral modes during parts of
each run, although it was not always the case that the direct and spectral images of a
given field were obtained during the same run.
 
All data reduction took place using the Image Reduction and Analysis Facility 
(IRAF\footnote{IRAF is distributed by the National Optical Astronomy Observatory, 
which is operated by the Association of Universities for Research in Astronomy, Inc., 
(AURA) under cooperative agreement with the National Science Foundation.}) software.  
A special package of IRAF-based routines that were written by members of the KISS team 
was used for most of the data analysis.  Full details of the observing procedures and data 
reduction methods are given in Paper I and KR1.  


\section{List 3 of the KPNO International Spectroscopic Survey}

\subsection{Selection Criteria}

The selection of the third red (H$\alpha$) list of ELG candidates was carried
out in precisely the same fashion as the first and second red lists (KR1 and KR2).  
Full details are
presented in Paper I and KR1.  To briefly summarize, we use our automated KISS software 
to evaluate the extracted objective-prism spectrum of each object located within a
survey field.  All objects with spectral features that rise more than five times the
local noise above the continuum level are flagged as potential ELGs.  This 5$\sigma$ 
threshold is the primary selection criterion of the survey, and was arrived at after
substantial testing during the early phases of the KISS project.  Following the
initial automated selection, all candidates are visually examined, and spurious
sources are removed from the sample.  Finally, the objective-prism images are
scanned visually for sources that might have been missed by the software.  These
tend to be objects where the emission line is redshifted to the red end of the
objective-prism spectrum, so that the software cannot detect continuum on both
sides of the line.  The combination of our automated selection process and our
careful visual checking helps to ensure a high degree of reliability that the KISS
ELG candidates are real, and that the sample is largely complete for all objects
with 5$\sigma$ emission lines.

As described in KR1, we also flag objects that have emission lines between
4$\sigma$ and 5$\sigma$ during our selection process.  These 4$\sigma$ detections
represent objects with somewhat weaker emission lines than the main KISS sample, but 
that are nonetheless valid ELG candidates.   However, these objects do not constitute a 
statistically complete sample in the same sense as the main ($>$ 5$\sigma$) list.  We 
report the 4--5$\sigma$ sources in a secondary list of ELG candidates (see Appendix), 
which should be thought of as a supplement to the main KISS catalog.  This list of 
``extra" (or KISSRx) objects likely includes a number of interesting sources. 

\subsection{The Survey}

The list of ELG candidates selected in the third red survey is presented 
in Table \ref{table:tab2}.  Because the survey data includes both spectral 
images and photometrically-calibrated direct images, we are 
able to include a great deal of useful information about each source, such  
as accurate photometry and astrometry and estimates of the redshift, 
emission-line flux and equivalent width.  Only the first page of the
table is printed here; the complete table is available in the electronic
version of this paper.

The contents of the survey table are as follows.  Column 1 gives a running 
number for each object in the survey with the designation KISSR $xxxx$, where 
KISSR stands for ``KISS red'' survey.  This is to distinguish it from the blue 
KISS survey (KB1).  The KR1 and KR2 survey lists included KISSR objects 1--2157, 
and here we present KISSR objects 2158--2418.  
Columns 2 and 3 give the object identification from the KISS database tables, 
where the first number indicates the survey field (F$xxxx$), and the second number 
is the identification number within the field table for that galaxy. This identifier is 
necessary for locating the KISS ELGs within the survey database tables. Columns 4 
and 5 list the right ascension and declination of each object (J2000). The formal
uncertainties in the coordinates are 0.25 arcsec in RA and 0.20 arcsec in
declination.  Column 6 gives the B magnitude, while column 7 lists the 
B$-$V color.  For brighter objects, the magnitude estimates typically have 
uncertainties of 0.05 mag, increasing to $\sim$0.10 mag at B = 20.
Paper I includes a complete discussion of the precision of both the astrometry 
and photometry of the KISS objects.  An estimate of the redshift of each
galaxy, based on its objective-prism spectrum, in given in column 8.
This estimate assumes that the emission line seen in the objective-prism 
spectrum is H$\alpha$.  Follow-up spectra for $>$1600 ELG candidates from the 
two red survey lists (KR1, KR2 and the current list) show that this assumption is correct 
in the vast majority of cases.  Only four ELGs in the current list that possess follow-up 
spectra (3\%) are high redshift objects where a different line (typically [\ion{O}{3}] and/or 
H$\beta$) appears in the objective-prism spectrum.  
The formal uncertainty in the redshift estimates is $\sigma_z$ = 0.0028 (see Section 
4.1.3).  Columns 9 and 10 list the emission-line flux (in units of 10$^{-16}$ erg/s/cm$^2$) 
and equivalent width (in \AA) measured from the objective-prism spectra.
The calibration of the fluxes is discussed in section 4.1.2.  These quantities should 
be taken as being representative estimates only.  A simple estimate of the reliability 
of each source, the quality flag (QFLAG), is given in column 11.  This quantity, 
assigned during the line measurement step of the data processing, is given the value 
of 1 for high quality sources, 2 for lower quality but still reliable objects, and 3 
for somewhat less reliable sources.  Column 12 gives alternate identifications for KISS 
ELGs which have been cataloged previously.  This is not an exhaustive cross-referencing, 
but focuses on previous objective-prism surveys which overlap part or all of the current 
survey area: Markarian (1967)  and Case (Pesch \& Sanduleak 1983). The Markarian survey 
overlaps both the Bo\"otes and Cetus fields, while the Case survey only overlaps the 
Bo\"otes field. Also included are objects in common with the {\it Uppsala General Catalogue 
of Galaxies} (UGC, Nilson 1973).

A total of 261 ELG candidates are included in this third list of H$\alpha$-selected 
KISS galaxies.  The total area covered by the third red survey strip is 19.65 deg$^2$, 
meaning that there are 13.3 KISS ELGs per square degree. For the first, second, and third red 
lists combined, the surface density is 16.4 galaxies deg$^{-2}$, and if the lower significance
KISSRx objects are included the density is 20.7 ELGs deg$^{-2}$.  This compares to the 
surface density of 0.1 galaxies deg$^{-2}$ from the Markarian (1967) survey, and 0.56 
galaxies per deg$^2$ from the H$\alpha$-selected UCM survey (Zamorano \etal 1994); the 
present survey is much deeper despite the redshift limit inherent in our detection method.  
It is interesting to note that the fraction of 4$\sigma$ -- 5$\sigma$ KISSRx ELG 
candidates is substantially higher for the sample presented here than for the first 
and second red survey areas.  As discussed below, we believe that this difference
is caused by the somewhat different noise characteristics of the CCD used for the current
survey.   For example, ELGs that would have been 5.0$\sigma$ objects when observed 
with the previous CCD might be detected as 4.8$\sigma$ sources in the current data.
The net effect would be to lower the number of objects in the main survey list and to
shift some of them into the KISSRx list.  If both the lower-significance KISSRx objects 
and the {\it bona fide} KISSR objects are combined, the surface density of ELG candidates 
is essentially constant for all three red survey lists.

\begin{figure*}[htp]
\vskip -0.5in
\epsfxsize=6.5in
\hskip 0.5in
\epsffile{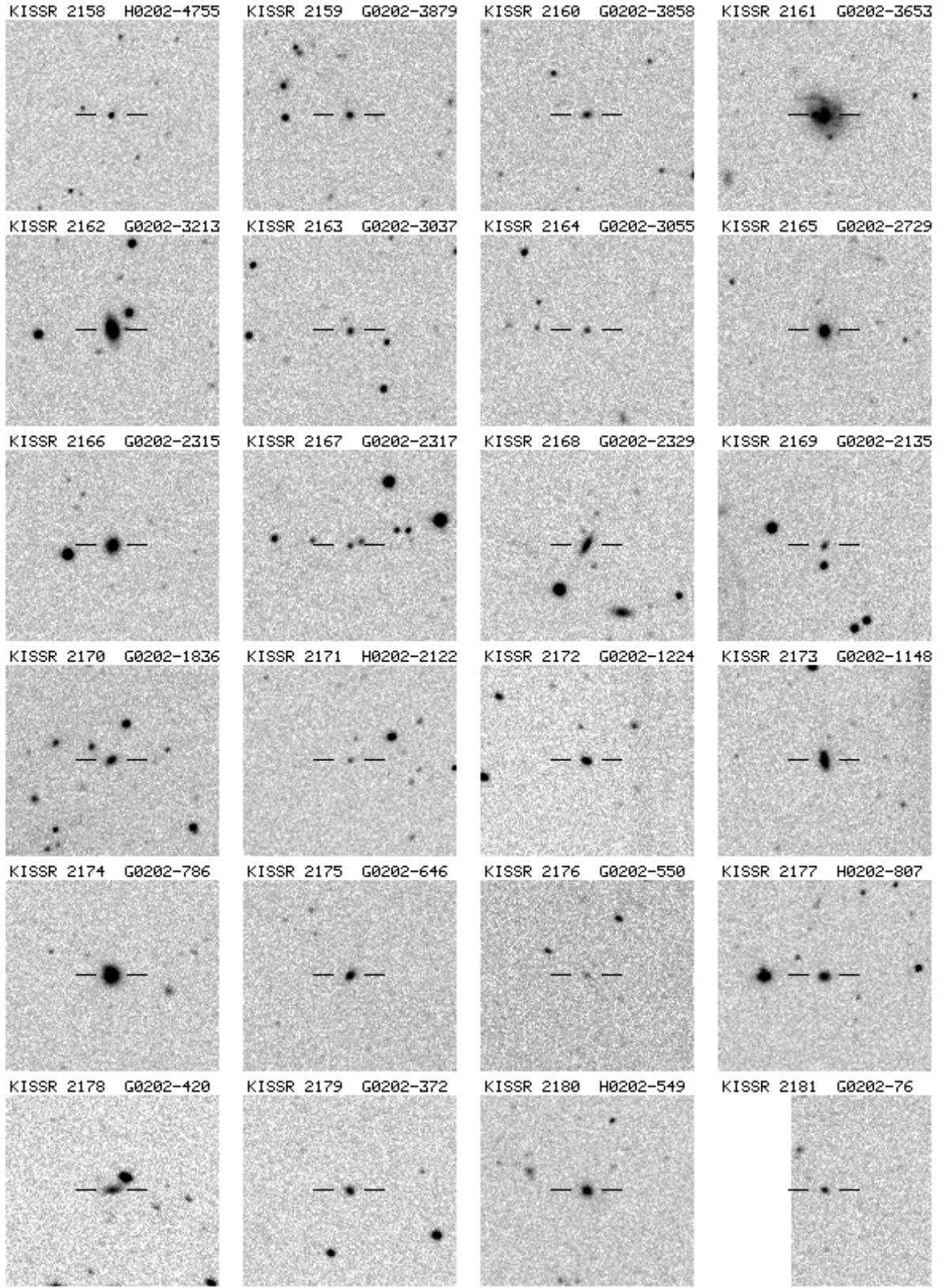}
\figcaption[find1.eps]{Example of finder charts for the KISS ELG candidates.
Each image is 3.2 $\times$ 2.9 arcmin, with N up, E left.  These finders are 
created from a composite of the B- and V-band direct images obtained as part 
of the survey.  In all cases the ELG candidate is located in the center of 
the image section displayed, and is indicated by the tick marks.
\label{fig:find1}}
\end{figure*}

\begin{figure*}[htp]
\vskip -0.5in
\epsfxsize=6.5in
\hskip 0.5in
\epsffile{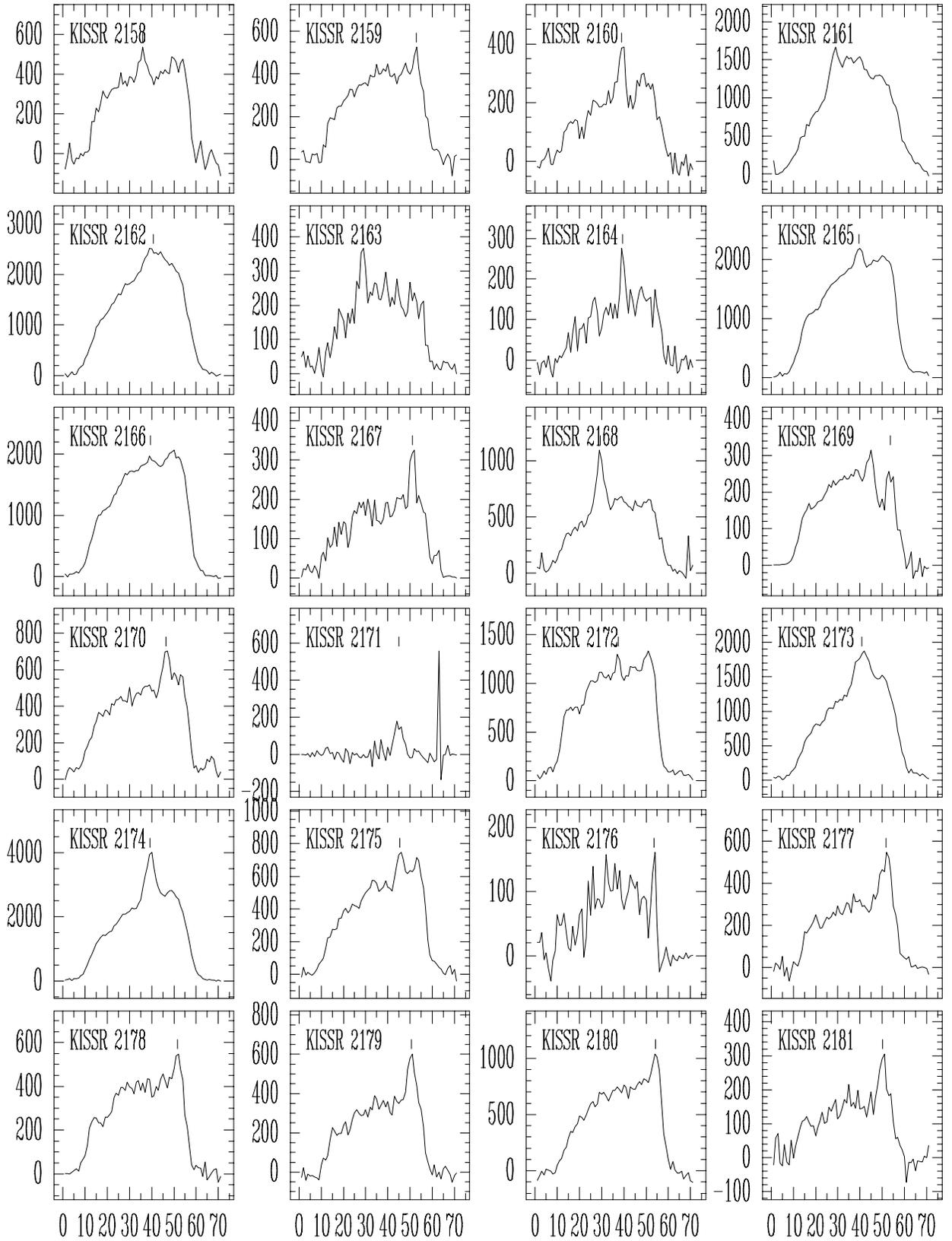}
\figcaption[spec1.eps]{Plots of the objective-prism spectra for 24 KISS ELG
candidates.  The spectral information displayed represents the extracted
spectra present in the KISS database tables.  The location of the putative 
emission line is indicated.
\label{fig:spec1}}
\end{figure*}

\begin{figure*}[htp]
\vskip -0.1in
\epsfxsize=5.0in
\hskip 1.0in
\epsffile{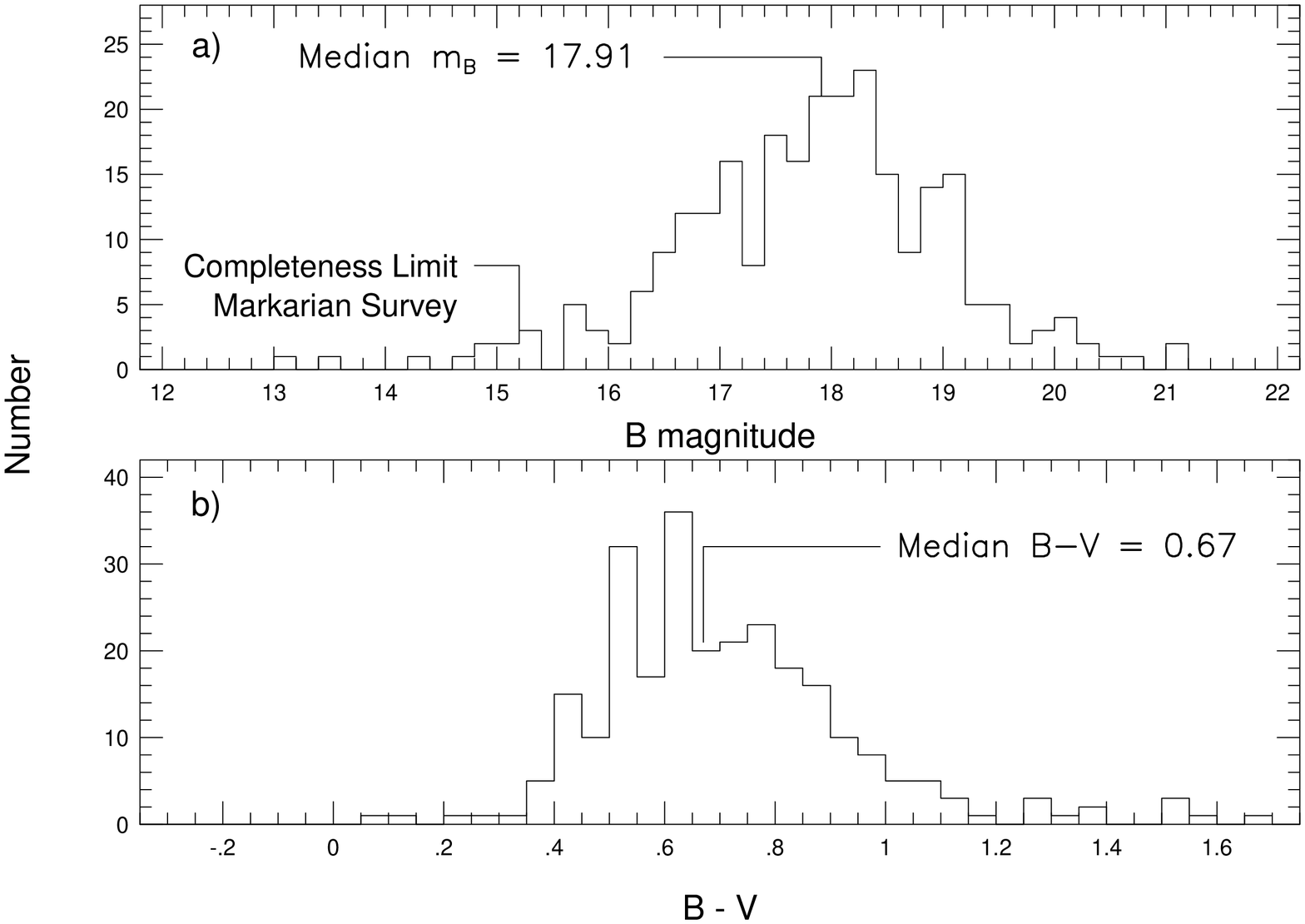}
\figcaption[fig3.eps]{(a) Distribution of B-band apparent magnitudes for the
261 ELG candidates in the third H$\alpha$-selected KISS survey list.  
The median brightness in the KISS sample is B = 17.91, with 4\% of the 
galaxies having B $>$ 20.  
Also indicated, for comparison, is the completeness limit of the Markarian 
survey. 
(b) Histogram of the B$-$V colors for the 261 ELG candidates.  
The median color of 0.67 is indicated. 
\label{fig:appmag}}
\end{figure*}

Of the 261 objects cataloged, 167 were assigned quality values of QFLAG = 1 (64\%), 74 
have QFLAG = 2 (28\%), and 20 have QFLAG = 3 (8\%). Based on our follow-up spectra
to date, 99\% (87 of 88) of the sources with QFLAG = 1 are {\it bona fide}
emission-line galaxies, compared to 83\% (24 of 29) with QFLAG = 2 and 78\% (7 of 9)
with QFLAG = 3.  Overall, 94\% of the objects with follow-up spectra are {\it bona fide} 
ELGs.  The properties of the KISS galaxy sample are described in the next section.

Figure~\ref{fig:find1} shows an example of the finder charts for the KISS ELGs.  
These are generated from the direct images obtained as part of the survey, and 
represent a composite of the B- and V-band images.  Figure~\ref{fig:spec1} 
displays the extracted spectra derived from the objective-prism images for the 
first 24 ELGs in Table 2.  Finder charts and spectral plots for all 261 objects 
in the current survey list, along with finder charts for the KISSRx objects, are 
available in the electronic version of this paper. 

A supplementary table containing an additional 158 ELG candidates with 4$\sigma$ to 
5$\sigma$ emission lines is included in the appendix of this paper (Table \ref{table:tab3}). 
These additional galaxies do not constitute a statistically complete sample, and should 
therefore be used with caution. However, there are likely many interesting objects contained 
in this supplementary list.  Hence, following the precedent established in KR1 and KR2, we 
list these objects in order to give a full accounting of the ELGs in the area surveyed.


\section{Properties of the KISS ELGs}

Due to the manner in which the survey is carried out, a great deal of observational data are
available for all of the KISS ELG candidates cataloged in the current paper.  This includes
accurate astrometry and B and V photometry for each source, as well as estimates of
the redshift, H$\alpha$ + [\ion{N}{2}] line fluxes and equivalent widths.  The combination
of these data allow us to acquire a fairly complete picture of the make-up of the KISS sample.
However, the quantities derived from the objective-prism spectra are inadequate for
detailed analyses.  First, the low resolution of the spectra limits the accuracy of the
redshifts  measured (see below).  Further, the combination of low resolution and limited
spectral coverage prevent us from using the survey data to ascertain the activity type
of the ELGs (e.g., AGN vs. star-forming).  Hence follow-up spectra obtained with a
higher dispersion spectrograph are required for a complete understanding of the
KISS ELGs.  Nonetheless, much can be gleamed about the survey constituents with
the data currently available.  We present an overview of the properties of our new
sample of KISS ELGs below.

\subsection{Observed Properties}

\subsubsection{Magnitude \& Color Distributions}

The B-band apparent magnitude distribution for the 261 KISS ELGs in the current survey 
list is shown in Figure~\ref{fig:appmag}a.  The median apparent B magnitude is 17.91.  This 
value is somewhat brighter than those of the KR1 and KR2 survey lists, which have median 
apparent magnitudes of B = 18.08 and 18.13, respectively.  However, it is clear that KISS 
still probes substantially deeper than previous objective-prism surveys: The median apparent 
magnitude for the H$\alpha$-selected UCM survey (P\'erez-Gonz\'alez \etal 2000) is 
B $\approx$ 16.1, and the [\ion{O}{3}]-selected Michigan (UM) survey (Salzer \etal 1989)
has a median apparent magnitude of B = 16.9.  Indicated in the figure is the completeness 
limit of the Markarian survey, B = 15.2 (Mazzarella \& Balzano 1989).  

\begin{figure*}[htp]
\vskip -0.1in
\epsfxsize=5.0in
\hskip 1.0in
\epsffile{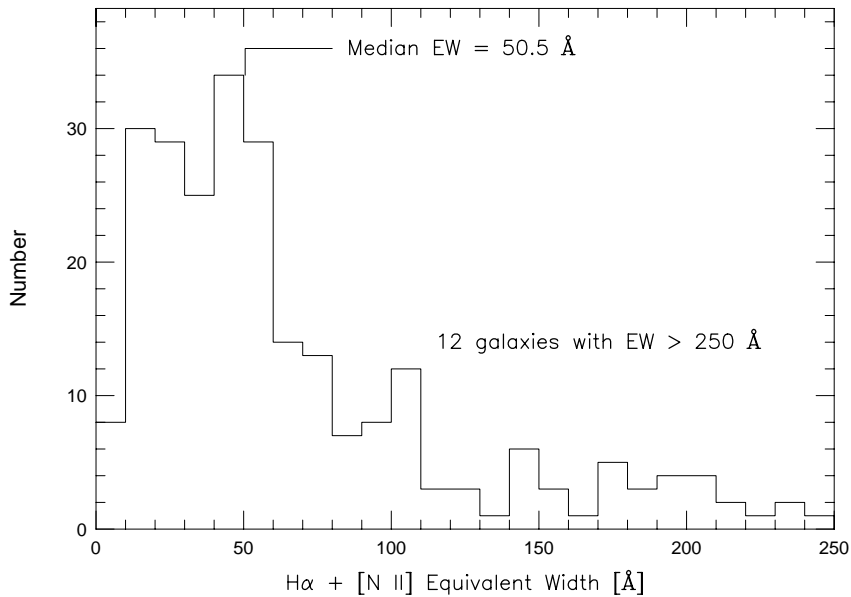}
\figcaption[fig4.eps]{Distribution of measured H$\alpha$ + [\ion{N}{2}]
equivalent widths for the KISS ELGs.  The median value of 50.5 \AA\ is 
indicated.  The measurements of equivalent widths from objective-prism 
spectra tend to show a large scatter when compared to equivalent widths 
from long-slit spectra, so these values should be taken only as estimates. 
The survey appears to detect most sources with EW(H$\alpha$+[\ion{N}{2}]) 
$>$ 30--40 \AA.
\label{fig:ew}}
\end{figure*}

The distribution of the B$-$V colors of the third red survey list is shown in Figure~\ref{fig:appmag}b. 
The median color is 0.67, which is identical to that of the first red survey list and very 
close to that 
of the KR2 survey list (B$-$V = 0.69).  This color is representative of an Sb galaxy (Roberts 
\& Haynes 1994). The UCM survey shows a similar color distribution, and has a median color of 
B$-$r = 0.71 (P\'erez-Gonz\'alez \etal 2000).  In contrast, the [\ion{O}{3}]-selected KB1 and UM
surveys have color distributions that are significantly shifted to the blue, with median B$-$V 
colors of 0.50 and 0.55, respectively (KB1; Salzer \etal 1989).  This is a selection effect 
caused by the
use of different emission lines for detection in the different surveys.  The H$\alpha$-selected
samples include a broader spectrum of ELGs, including many more luminous star-forming
galaxies and LINERS which tend to be dominated by older, redder stellar populations.  
In addition, they are able to detect galaxies with higher levels of intrinsic reddening.  
The [\ion{O}{3}]-selected 
samples are dominated by lower luminosity, lower metallicity galaxies which are dominated by
younger stellar populations and have lower levels of internal absorption and reddening.
While the H$\alpha$-selected surveys tend to include both types of ELGs, they are dominated
by the more luminous galaxies.  In contrast, the blue-selected surveys tend to not select the
redder galaxies at all.

\subsubsection{Line Strength Distributions and Survey Completeness}

The distribution of equivalent widths (EWs) for the third red survey list is shown in 
Figure~\ref{fig:ew}. We assume that the line we measure in the survey spectra is H$\alpha$ 
blended with the [\ion{N}{2}]$\lambda\lambda$6584,6548 lines. Based on follow-up observations 
obtained to date, we know this assumption to be true for the vast majority of red survey 
objects. The three lines are blended at the resolution of the objective-prism spectra.
The [\ion{S}{2}]$\lambda\lambda$6731,6717 doublet is well resolved from the blended 
H$\alpha$ + [\ion{N}{2}] lines, and is often seen in survey spectra from strong-lined objects.
The EW distribution peaks in the 40 -- 50 \AA\ bin, which indicates that KISS is fairly 
complete for objects with equivalent widths greater than $\sim 50$ \AA. 
The median equivalent width of H$\alpha$ + [\ion{N}{2}] is 50.5 \AA, with the majority of 
ELGs having equivalent widths of less than 100 \AA. This median EW is approximately 25\% higher 
for this sample than for the two previous red survey lists. The noise level in the
third red survey list data is slightly higher than for the first and second red survey list.
We attribute this to the use of a different CCD for the newer sample, which had somewhat
worse noise characteristics than the previous CCD.  This shift in noise characteristics 
results in a selection of 5$\sigma$ ELG candidates that have relatively stronger lines. As
we mention above the fraction of 4$\sigma$ -- 5$\sigma$ objects is higher for the survey 
list presented here than for the two previous red survey lists, which is what we expect due 
to a higher noise level.

\begin{figure*}[htp]
\vskip -0.1in
\epsfxsize=5.0in
\hskip 1.0in
\epsffile{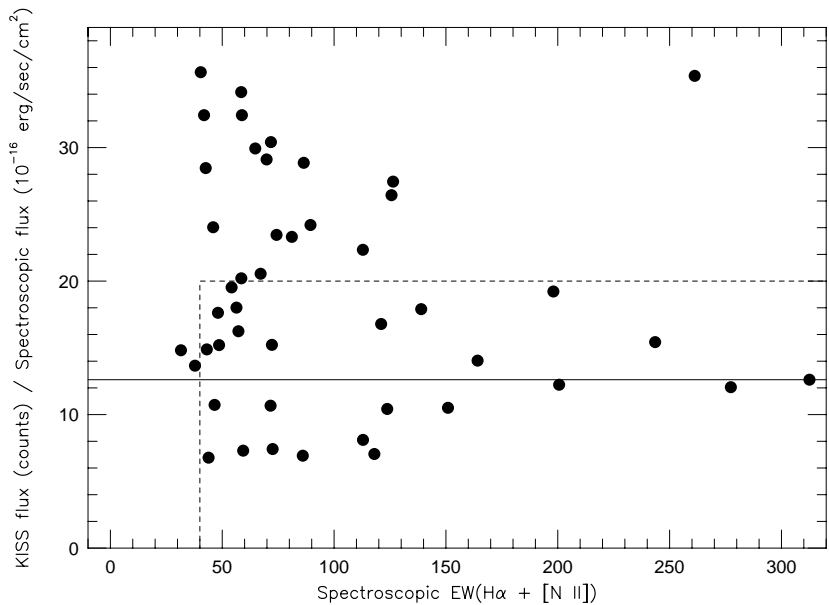}
\figcaption[fig5.eps]{Plot of the ratio of objective-prism flux (in counts) 
to spectroscopic flux {\it versus} H$\alpha$~+~[\ion{N}{2}] equivalent width 
measured 
from the follow-up spectra. The plotted data are all from the same observing 
run (KPNO 2.1-m, 2004). 
The solid line indicates the median ratio.  
The dashed lines show the criteria we applied to select the calibration sample. 
One galaxy with flux ratio $>$~40 falls above the diagram, and another galaxy 
with EW $>$~320~\AA\ lies off the diagram to the right.
\label{fig:lfluxcal}}
\end{figure*}

The calibration of the flux scale is a two-step process. The objective-prism spectra 
for each field are first corrected for throughput variations and atmospheric extinction. 
This places all line fluxes on the same {\it relative} flux scale.
The fluxes are then calibrated on an absolute scale, using information obtained from 
the follow-up spectra. From a sample of 126 follow-up observations, we selected 49
objects that were classified as star-forming galaxies, had spectral quality
$Q = 1$ or 2, and had been observed with a long-slit spectrograph under photometric 
conditions. All spectra were obtained during the same observing run, and are of galaxies
located in the Bo\"otes field.
Since the fluxes measured from the objective-prism spectra are a combination of the 
H$\alpha$ and [\ion{N}{2}] lines, we use the fluxes from our slit spectra for the sum 
of these three lines. Figure~\ref{fig:lfluxcal} shows a plot of the ratio of 
objective-prism flux (in counts) to spectroscopic flux {\it versus} the equivalent 
width measured from the follow-up spectra. The majority of the emission-line flux 
from a point source was included in the follow-up spectra, since the slit width used 
was 2.0\arcsec. Some galaxies, however, have a larger angular extent and the emission 
lines originate in an extended region. Since our long-slit measurements do not include 
all the H$\alpha$ emission from these sources, they tend to have large flux ratios.
We restricted the  calibration sample to those galaxies with an 
objective-prism-to-spectroscopic flux ratio of less than 20 and an equivalent width 
greater than 40 \AA, which left 25 galaxies. 
The emission regions of these galaxies are essentially point sources.
The median flux ratio of the calibration sample is 12.61; the mean is 12.92 with a 
standard deviation of 4.19 and an error in the mean of 0.84. We adopted the reciprocal 
of the median value as our calibration value, or $0.0793 \times 10^{-16}$ ergs/s/cm$^2$ 
per count. 

\begin{figure*}[htp]
\vskip -0.1in
\epsfxsize=5.0in
\hskip 1.0in
\epsffile{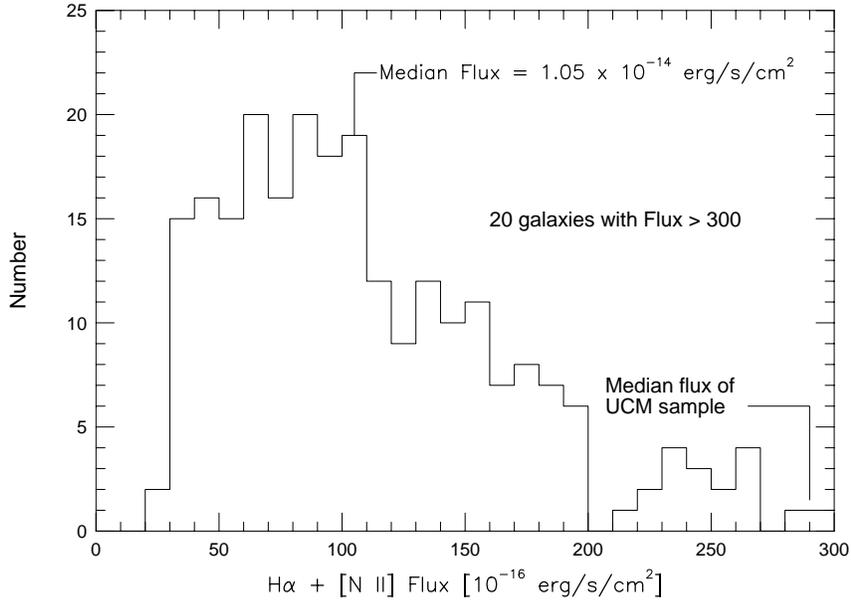}
\figcaption[fig6.eps]{Distribution of H$\alpha$ + [\ion{N}{2}] line fluxes
for the 261 KISS ELGs included in the current survey list.  The median flux
level of both the KISS and UCM samples is indicated.
\label{fig:lflux}}
\end{figure*}

The calibration value is applied to the measured objective-prism line fluxes to convert 
their instrumental fluxes (in counts) to calibrated fluxes (in ergs/s/cm$^2$). 
In Figure~\ref{fig:lflux} we show the distribution of observed H$\alpha$ + [\ion{N}{2}] 
line flux values for the 261 KISS ELGs. The median value is 
$1.05 \times 10^{-14}$ ergs/s/cm$^2$, which is $\sim 30\%$ higher than the values found 
for the the first two red survey lists, suggesting that the data used for the current 
survey is less sensitive that those used for the first and second red survey lists. 
As we mention above, this difference is likely due to the slightly higher noise level in 
the KR3 data. However, the median line flux of the third red survey list is substantially 
fainter than that of the UCM sample, which is $2.9 \times 10^{-14}$ ergs/s/cm$^2$ 
(based on follow-up spectra of Gallego \etal 1996). 

\begin{figure*}[htp]
\vskip -0.0in
\epsfxsize=5.0in
\hskip 1.0in
\epsffile{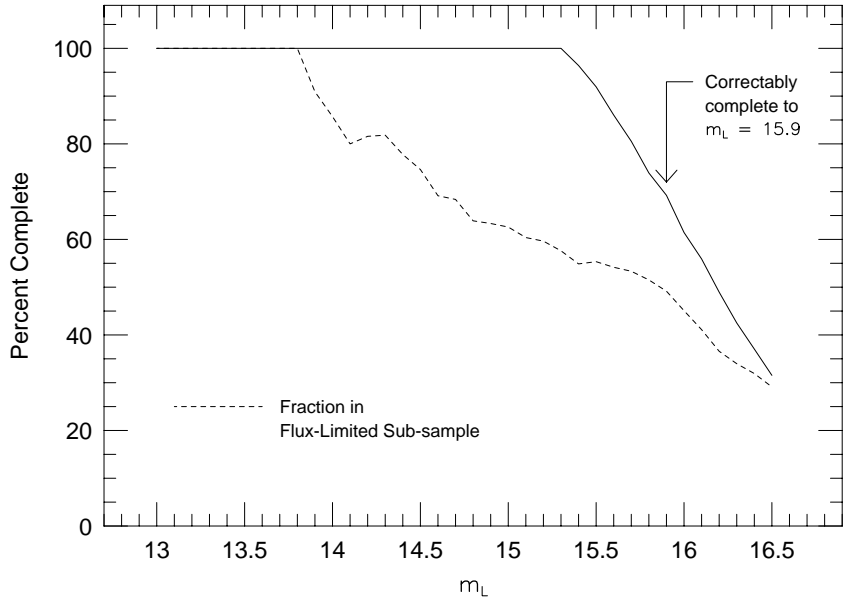}
\figcaption[complete_kr3.eps]{Plot of the completeness percentage as a function of 
m$_L$ for the current sample (solid line).  The catalog is 100\% complete to
m$_L$ = 15.3, and is ``correctably complete" to m$_L$ = 15.9. 
The dashed line shows the fraction of the sample contained in the 
flux-limited sub-sample as a function of m$_L$.  At the completeness 
limit, roughly half of the KISS ELGs are in the flux-limited portion of 
the sample. \label{fig:complete}}
\end{figure*}

As mentioned earlier, one of the strengths of KISS is that the selection function
and completeness limit can be derived using the survey data directly, rather
than relying on secondary information (e.g., line strengths measured from
follow-up spectra).  The calibrated objective-prism line fluxes are used to 
determine the completeness limit of the survey, following the procedure described 
in Gronwall \etal (2005, in preparation).  Briefly, we convert the line fluxes into 
pseudo-magnitudes --  the line magnitude m$_L$, and then apply a V/V$_{max}$
analysis (e.g., Schmidt 1968, Huchra \& Sargent 1973) to the complete sample of
261 galaxies.  The results are presented in Table~\ref{table:comptab}.  Column (1)
lists m$_{comp}$, the value of m$_L$ for which $\langle V/V_{max}\rangle$
is being computed.  Column (2) lists the total number of ELGs brighter 
than that m$_L$ level, while columns (3) and (4) give the numbers of objects
in the volume-limited and flux-limited subsamples, respectively (see below).  Note 
that some objects may start out in the flux-limited sample at brighter values of
m$_{comp}$, then move into the volume-limited sample at fainter values of
m$_{comp}$.  Column (5) lists the mean V/V$_{max}$ for the flux-limited 
subsample.  Column (6) shows the number of galaxies that need to be added 
to the sample at each m$_{comp}$ level to maintain $\langle V/V_{max}\rangle 
= 0.5$, and column (7) lists the cumulative number of galaxies added at all 
magnitudes brighter than the given magnitude level to maintain
$\langle V/V_{max}\rangle$.  Column (8) shows the percentage of objects that
are in the flux-limited subsample, which decreases continuously as m$_L$ becomes
fainter.  Column (9) lists the completeness percentage of the flux-limited 
subsample as a function of m$_L$.   These latter two quantities are plotted in 
Figure~\ref{fig:complete}.

\begin{figure*}[htp]
\vskip -0.1in
\epsfxsize=5.0in
\hskip 1.0in
\epsffile{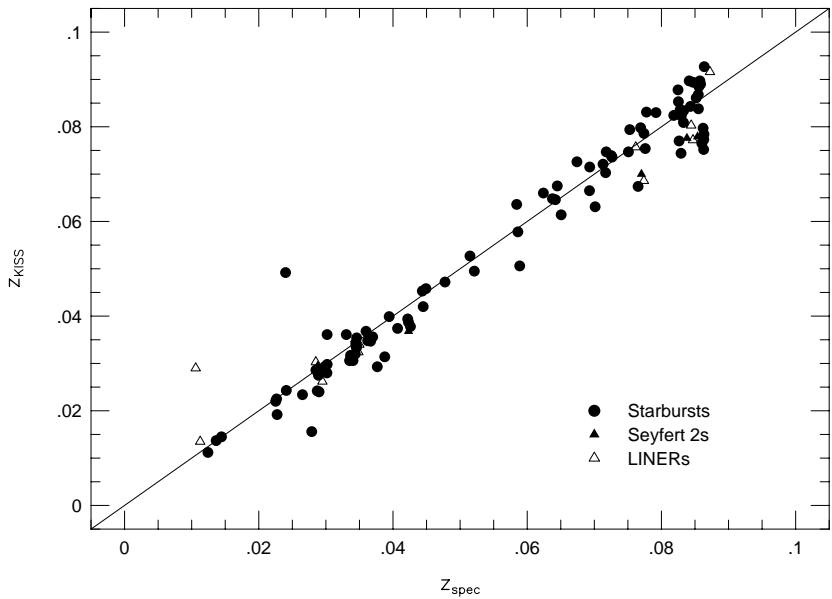}
\figcaption[zcomp.eps]{Comparison between objective-prism redshifts (z$_{KISS}$) 
obtained from our survey data and slit-spectra redshifts (z$_{spec}$) obtained from
follow-up spectra.  The solid line denotes z$_{KISS}$ = z$_{spec}$.   The objective-prism
redshifts provide reasonable estimates of the true redshifts over the full range covered
by the survey.  The formal uncertainty in z$_{KISS}$ is 0.0032 (950 \kms). \label{fig:zcompare}}
\end{figure*}

The interpretation of the results of the V/V$_{max}$ test follows exactly the discussion 
found in Gronwall \etal (2005, in preparation) for the KR1 sample.  Rather than repeating
that discussion here, we will simply summarize the main results.  It is important to realize 
that because of the redshift limit imposed by the survey filter, objects in the sample can 
be either line-flux-limited or volume-limited objects, depending on the strength of their 
H$\alpha$+[\ion{N}{2}] emission and their redshift.  Objects with sufficiently strong lines
will have values of V$_{max}$ that exceed the effective volume of the survey set by the
redshift limit.  Such objects are volume limited.  As the limiting line flux (parameterized 
by m$_L$) decreases, a given object may actually switch from being in the flux-limited 
category to the volume-limited category.  As seen in the table, for faint limiting line 
fluxes (fainter m$_L$) the majority of the KISS ELGs are in the volume-limited subsample.  
This is illustrated by the dashed line in Figure~\ref{fig:complete}.  We see that the KISS 
sample is 100\% complete to m$_L$ = 15.3, which is very similar to the results for the KR1 
and KR2 samples (Gronwall \etal 2004, 2005).  This completeness limit includes 184 KISS ELGs, 
or 70..5\% of the full sample.  As is often done, one can construct a ``correctably complete" 
sample by extending the line-flux limit down to even lower values.  For example, at m$_L$ = 15.9, 
the sample is still 69.2\% complete, but now includes 91.2\% of the sample.

\subsubsection{Redshift Comparison and Distributions}

We also derive redshifts from the objective-prism spectra. For objects with 
follow-up observations, we can compare the survey redshifts to the redshifts 
derived from the long-slit spectra (Figure~\ref{fig:zcompare}). In general the 
agreement between $z_{KISS}$ (objective-prism redshift) and $z_{spec}$ (follow-up 
redshift) is excellent. Only four objects deviate substantially from the equality 
line. Two of these are active galaxies at $z > 0.35$ which are not shown in the 
diagram. They were detected due to their [\ion{O}{3}]$\lambda$5007 and H$\beta$ 
lines, respectively.  The remaining two 
objects are KISSR 2336 and KISSR 2320; two relatively large, well-resolved disk 
galaxies that both have emission regions which are offset from the center of the galaxy. 
Because the dispersion of the objective-prism spectra is in the north-south direction, 
a spectrum of an object that has an emission region spatially offset north or south 
of the center will yield an incorrect estimate of the redshift of the emission line. 
Only a small minority of KISS objects are affected by this.

For the first red survey list (KR1) the survey redshifts above $z_{KISS} = 0.07$ 
showed a systematic offset from the redshifts determined from follow-up spectroscopy.
The reason for the offset is that as the H$\alpha$ + [\ion{N}{2}] line in the 
objective-prism spectrum begins to shift out of the survey bandpass, only the 
lower-redshift portion of the line is detected. A correction was applied to the KR1 
survey redshifts, as described in Paper I. The objects plotted in Figure~\ref{fig:zcompare}
do not display this offset, and no correction was applied to the survey redshifts. 
The second red survey list (KR2) also did not show any systematic offset. The reason 
for the difference is probably the better pixel scale of the CCD used for 
the second and third red survey lists. When we calculate the RMS scatter of $z_{KISS}$ 
about the equality line we use only objects with $z_{spec} < 0.07$, as was done for
KR1.  We exclude  the four most deviant objects that we described in the previous 
paragraph. The resulting RMS scatter is 0.0032 (950 km/s), which is marginally higher 
than the value found for the first two red survey lists (0.0028, or 840 km/s).  

\begin{figure*}[htp]
\vskip -0.1in
\epsfxsize=5.0in
\hskip 1.0in
\epsffile{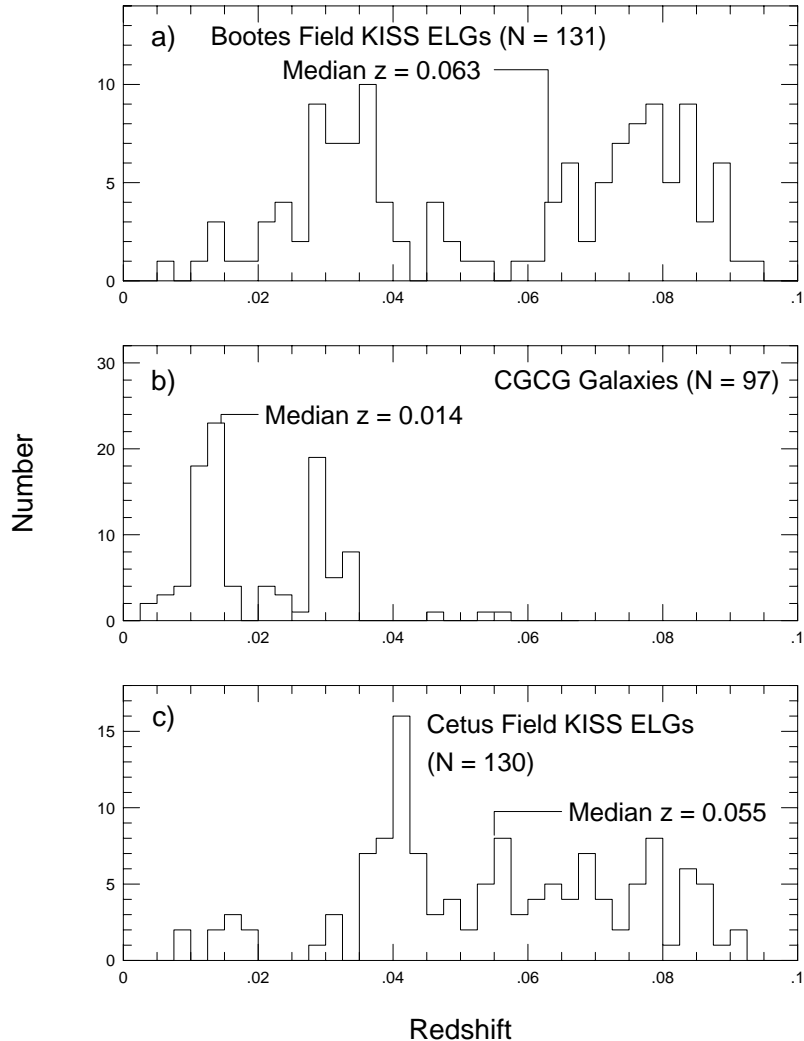}
\figcaption[fig9.eps]{Histograms showing the distribution of redshift for the 261 
H$\alpha$-selected KISS ELGs in (a) the Bo\"otes field and (c) the Cetus field. 
In $(b)$ we show the distribution of 97 ``normal" galaxies from the 
CGCG that are located in the area of the Bo\"otes field.
The median redshift is indicated in all plots. 
The deficit of ELGs between z = 0.04 and 0.06 seen in (a) is due to the 
Bo\"otes void.  \label{fig:zhist}}
\end{figure*}

The distribution of the objective-prism redshifts is shown in Figure~\ref{fig:zhist}. 
The Bo\"otes and Cetus fields are shown in separate panels. The middle panel shows 
the redshift distribution for a comparison sample of galaxies from Zwicky \etal 
(1961 -- 1968; hereafter CGCG). The redshifts for the 97 CGCG galaxies are taken 
from Falco \etal (1999). Since the surface density of the CGCG catalog is fairly 
low, we included objects located in a region of $9 \times 9$ degrees, centered at 
the location of the Bo\"otes field. The CGCG catalog does not extend far enough south 
to overlap the Cetus field, hence the comparison sample applies only to the upper
(Bo\"otes) redshift sample.

The Bo\"otes void (Kirshner \etal 1981) is clearly visible between $z = 0.04$ 
and $z = 0.06$ in Figure~\ref{fig:zhist}a.  
Even thought the NDWFS field is located far south of the nominal void
center, the impact of the void is unmistakable in the redshift distribution.
There is a significant density enhancement seen at redshifts between
0.0275 and 0.0375, just in front of the void.  This is most likely 
associated with the Hercules supercluster.  A modest peak in the redshift
distribution of the CGCG sample is also seen in this redshift range.
The latter becomes quite sparse beyond this redshift.  The CGCG sample
also shows a strong peak at z = 0.0125 which is less prominent in the
KISS distribution.   Of the 41 CGCG galaxies that constitute this peak, only 
four are found inside the Bo\"otes field survey area. The galaxies in this 
peak appear to fall along a large-scale structure feature that falls mostly 
outside the NDWFS area.  Beyond the Bo\"otes void the KISS sample displays
a fairly flat redshift distribution out to z~$\sim$~0.085, after which
it begins to drop off.  The distribution drops to zero at z = 0.095, because 
KISS cannot detect galaxies via the H$\alpha$ line beyond this distance since 
it redshifts completely out of the survey filter at this point.  The flat
distribution and drop-off between 0.085 and 0.095 are characteristic of
what is seen with the KR1 and KR2 samples as well.

\begin{figure*}[htp]
\vskip -0.1in
\epsfxsize=5.0in
\hskip 1.0in
\epsffile{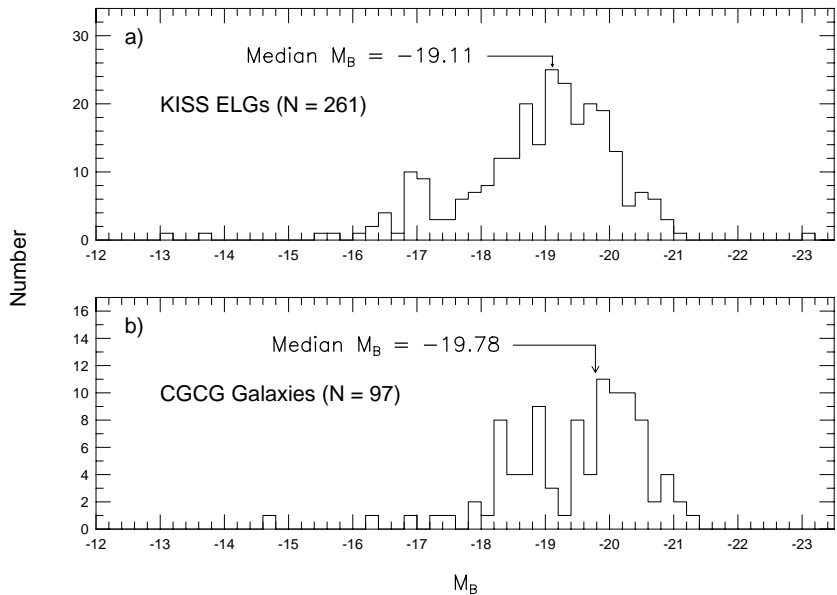}
\figcaption[fig8.eps]{Histograms showing the distribution of blue absolute 
magnitude for $(a)$ the H$\alpha$-selected KISS ELGs from the two survey fields 
and $(b)$ the 97 ``normal" galaxies from the CGCG that are located in the Bo\"otes field 
area of the sky.  
The median luminosity of each sample is indicated.  
The KISS ELG sample is made up of predominantly intermediate- and 
lower-luminosity galaxies, making this line-selected sample particularly 
powerful for studying dwarf galaxies.  \label{fig:absmag}}
\end{figure*}

The redshift distribution for the Cetus field (Figure~\ref{fig:zhist}c)
is dominated by a low density region at low redshifts that reaches out to
z = 0.035.  This is part of the large void that dominates the foreground
in the south Galactic cap.  There is no hint of the Pisces-Perseus
supercluster at z~$\approx$~0.020; the Cetus field is far enough south
to miss the supercluster.  There is a strong density enhancement at
z = 0.040.  Beyond this point, the KISS redshift distribution
is again fairly flat out the the point where the survey filter starts
to exclude the H$\alpha$ line.  Note that the location of the Cetus
field is such that only very sparse redshift information of the ``normal"
galaxies exists.  Hence, we do not have a suitable comparison sample as
we do with the Bo\"otes field.

The median redshifts of the two NDWFS KISS samples are quite similar to the
values found for KR1 (median z = 0.063) and KR2 (median z = 0.061).

\subsection{Luminosity Distributions}

The availability of both an accurate apparent magnitude and a redshift
estimate allows us to compute the absolute magnitude for each source.
Using the values listed in Table 2 for the B magnitude and
objective-prism redshift, we compute M$_B$ for all ELG candidates in
the current list.  H$_o$ = 75 km/s/Mpc is adopted, and a correction for
Galactic absorption is applied using the values for A$_B$ compiled by
Schlegel \etal (1998).  In both fields the Galactic absorption is small,
with typical values of 0.04 -- 0.06 mags in the Bo\"otes field, and 0.09 
-- 0.11 mags in the Cetus region.   An explicit assumption made is that the 
line seen in the objective-prism spectrum is in fact H$\alpha$.  Previous
observations of KR1 and KR2 ELGs suggest that for roughly 2.5\% of the 
KISS ELG candidates the line seen in the objective-prism spectrum is 
some other line (usually [\ion{O}{3}]$\lambda$5007).  Hence we might 
expect 6-7 of the ELG candidates in Table 2 to fall in this category.  

A histogram showing the distribution of M$_B$ for the KISS ELGs in the
current survey list is shown in Figure 10a.  For comparison, in Figure 10b
we plot the absolute magnitude distribution for the 97 CGCG galaxies 
used in the redshift distribution comparison in the previous subsection.
Note that for this presentation we have combined the two portions of the
KISS NDWFS sample into a single histogram.  The median absolute magnitudes
of the KISS and CGCG galaxies differ significantly:  While the CGCG has
a median M$_B$ = $-$19.78 (i.e., close to M$^*$), the KISS ELGs have a
median absolute magnitude fully 2/3rds of a magnitude fainter.  This is
consistent with our previous survey lists.  That is, KISS is especially
sensitive to intermediate and low luminosity galaxies when compared to
a magnitude-limited sample like the CGCG.

Despite the fact that the current sample of KISS ELGs have systematically 
lower luminosities than the CGCG galaxies located in the same area of the 
sky, they are, on average, {\it higher} in luminosity than either of the
previous two H$\alpha$-selected KISS lists.  The median M$_B$ values for
KR1 and KR2 are $-$18.96 and $-$18.64, respectively.  The apparent reason
for the differences between the three lists is the relative paucity of
lower redshift galaxies in the current survey.  As mentioned above, 
the Cetus and Bo\"otes fields of the NDWFS exhibit very low numbers of
galaxies at redshifts below 0.035 and 0.025, respectively.  This is
precisely where KISS is most sensitive to dwarf ELGs.  To be sure, there
are still plenty of low luminosity galaxies present in the current survey
list.  However, the fraction of the survey with M$_B$ $>$ $-$18 is less
than in both KR1 and KR2.

At the high luminosity end, the KISS ELGs appear to be deficient in
galaxies with luminosities above M$^*$ (M$_B$ $<$ $-$20).  While the
KISS ELGs cover the same absolute magnitude range as the magnitude-limited
CGCG galaxies, the proportion of higher luminosity galaxies is much lower
among the KISS galaxies.  This most likely occurs for two reasons.  First,
the KISS sample is redshift limited at high luminosities, meaning that it
does not probe arbitrarily large volumes of space for the highest luminosity
objects as magnitude-limited samples do.  Second, the detection of any
galaxy by KISS requires that the emission line observed be bright enough
to stand out against the stellar continuum of the galaxy in the 
objective-prism spectrum.  That is, there must be a minimum contrast
between the continuum and the line.  This is effectively an equivalent
width limit.  For luminous galaxies, a larger star-forming event or a
stronger level of AGN activity is needed for the emission lines to exceed 
this implicit equivalent width threshold.  Since these intense activity
levels are fairly rare, there are fewer detected ELGs among the higher
luminosity host galaxies.  The combination of these two effects means
that there are fewer KISS ELGs at luminosities above M$^*$.

\subsection{Comparison with Previous Surveys}

Table \ref{table:tab2} lists cross-references for KISS ELGs which are also cataloged 
in previous photographic surveys for active and star-forming galaxies, and we note the 
UGC numbers for the objects that are listed in the {\it Uppsala General Catalogue of 
Galaxies} (Nilson, 1973). The first red survey area overlapped with four major active 
galaxy surveys: Markarian (1967), Case (Pesch \& Sanduleak 1983), Wasilewski (1983), and 
UCM (Zamorano \etal 1994).  The third red survey area, like KR2, overlaps with only the 
Markarian and Case surveys. 

The Markarian survey overlaps both the Bo\"otes and Cetus fields. There are, however, no 
Markarian galaxies in either field.  This is not too surprising, since the surface density 
of Markarian galaxies is small (0.1 per sq. deg.).  The Case survey overlaps only with the 
Bo\"otes field, and there are 18 Case objects in this area.  However, two of them lie just
outside the area covered by the KISS objective-prism images. They are both included in the
KISS direct images, but as we mention in \S2, the spectral and direct images do not always 
cover the exact same area. 
Two additional Case objects (459 and 460) are emission regions within the same galaxy,
and we choose to count them as one object for the purpose of the comparison with KISS.
Of the resulting 15 Case objects, 13 (87\%) are recovered by KISS in the main survey 
(i.e., Table 2). This fraction of recovered objects is somewhat higher than what was found 
for the first two red surveys (73\% and 72\% respectively). Both of the two Case galaxies 
that KISS does not recover are listed as color-selected in the Case survey papers.
Neither one is listed in the secondary KISS survey list with 4$\sigma$ to 5$\sigma$ ELG
candidates. A large fraction of Case galaxies have H$\alpha$ lines with equivalent widths 
less than 30 \AA \ (Salzer \etal 1995), and KISS is not as sensitive to this type of object.

The UGC catalog overlaps with the Bo\"otes field, and there are eight UGC galaxies 
in this area. Four of them are also KISS galaxies. Weak emission lines appear to be 
present in the objective-prism spectra of the remaining four galaxies, but only at 
the $\sim$3$\sigma$ level.


\section{Discussion \& Summary}

We present the third list of H$\alpha$-selected emission-line galaxy candidates
(and fourth list overall) from the KPNO International Spectroscopic Survey (KISS).
All data presented here were obtained with the 0.61-m Burrell Schmidt telescope.
KISS is an objective-prism survey, but differs from older such surveys by virtue
of the fact that it utilizes a CCD as the detector.  While we sacrifice areal coverage
relative to classical photographic surveys, we benefit from the enormous gain in
sensitivity that CCDs provide over plates.   We readily detect strong-lined ELGs as 
faint as B = 21.  In addition, the pan-chromatic nature of CCDs allows us greater
wavelength agility compared to photographic surveys.  Even with the use of our
survey filter, which restricts the detection of ELGs to z $<$ 0.095, we are sensitive 
to a broader range of galaxian redshifts than the older photographic objective-prism 
surveys (Paper I).  The combination of higher sensitivity, 
lower noise, and larger volumes surveyed yield huge improvements in the depth of the 
resulting survey.  With the KISSRx objects included, KISS finds $>$200 times more AGN 
and starburst galaxy candidates per unit area than did the Markarian (1967) survey, 
and $\sim$37 times more than the UCM survey (Zamorano \etal 1994).

The current installment of KISS includes 261 ELG candidates selected from 20 red 
survey fields covering a total of 19.65 deg$^2$.  This yields a surface density of 13.3
galaxies per deg$^2$.   We are sensitive to the H$\alpha$ emission line with redshifts 
up to $\sim$0.10.   The survey fields presented here are located at 
RA = 14$^h$~30$^m$, $\delta$ = 34$\arcdeg$~30$\arcmin$ (B1950), and at 
RA = 2$^h$~7.5$^m$, $\delta$ = -4$\arcdeg$~44$\arcmin$.
These fields were chosen to coincide with the location of the NOAO Deep Wide-Field 
Survey (Jannuzi \& Dey 1999).  
For each object in the catalog we tabulate accurate equatorial coordinates, B \& V 
photometry, and estimates of the redshift and line strength measured from the 
objective-prism spectra. We also provide finder charts and extracted spectral 
plots for all objects.  In addition to the main survey list, we include a 
supplementary list of 158 ELG candidates with weaker (lower significance) 
emission lines.  

This newest list brings the total of H$\alpha$-selected KISSR ELGs to 
2418 objects present in three survey regions.  In addition, we have cataloged another 638 
``extra" KISSRx candidates that are detected in the survey with a lower significance level,  
The total number of cataloged ELGs is 3056, contained in a survey area of just 147.6 sq.
deg. The overall surface density of KISS ELGs is thus 20.7 per sq. deg.

One of the advantages of our survey method is the large amount of basic data
that we acquire for each object.  This in turn allows us to parameterize the 
constituents of the survey and to develop a fairly complete picture of the overall
sample without the need for extensive follow-up observations.  We present
an overview of the survey properties for the current list of ELG candidates.
The median apparent magnitude of the current sample is B~=~17.91. This is 
somewhat brighter than the values found for KR1 and KR2 (B = 18.08 and 18.13, 
respectively), but it is still substantially fainter than previous ELG surveys.  
Objects fainter than B~=~20 are routinely cataloged.  
Line strengths measured from the objective-prism spectra show that KISS is 
sensitive to objects with H$\alpha$ + [\ion{N}{2}] equivalent widths of less than 
20 \AA, and that most objects with EW $>$ 40 \AA\ are detected. The median 
emission-line flux of the KISS sample is nearly three times lower than that of 
the UCM survey (Gallego \etal 1996).  The luminosity distribution of the KISS ELGs 
is heavily weighted toward intermediate- and low-luminosity galaxies, although we
are still sensitive to luminous AGN and starbursting galaxies.  The median absolute 
magnitude of M$_B$ = $-$19.11 underscores the fact that strong-lined galaxies of 
the type cataloged by KISS tend to be less luminous than the types of objects found 
in more traditional magnitude-limited samples.

Despite the fact that one can learn a great deal about each KISS ELG
from the survey data alone, it is still necessary to obtain higher dispersion
follow-up spectra in order to arrive at a more complete understanding of each object.
For example, due to the low-dispersion nature of the objective-prism spectra it is not 
possible to distinguish between AGN and star-formation activity in the KISS galaxies.
Further, the redshifts derived from the KISS spectral data are too coarse to be used 
in detailed spatial distribution studies. We 
have obtained follow-up spectra for 100\% of KISS ELGs in the KR1 and KB1 survey 
lists, and we are in the process of obtaining spectra for the objects in the remaining 
survey lists (KR2 and the current list. These follow-up spectra will allow us to better 
assess the nature of the individual galaxies, which in turn enables the sample to be 
used for a wide variety of science applications, many of which are outlined in Paper I.  
Examples include a series of multi-wavelength studies of the properties of KISS ELGs 
in the radio (Van Duyne \etal 2004) and X-rays (Stevenson \etal 2002, Datta \etal 2006, 
in preparation), plus studies currently underway in the mid- and far-IR (IRAS and 
Spitzer) and near-IR (2MASS), as well as ongoing studies of the metal abundances in 
KISS star-forming galaxies (e.g., Lee \etal 2004, Salzer \etal 2005).

\acknowledgments

We gratefully acknowledge financial support for the KISS project through
NSF Presidential Faculty Award to JJS (NSF-AST-9553020), which was instrumental
in initiating the original international collaboration, as well as continued support via
NSF-AST-0071114 and NSF-AST-0307766.  CG also acknowledges support from 
NSF-AST-0137927.  
We thank the numerous colleagues with whom we have 
discussed the KISS project over the past several years, including Jes\'us Gallego, 
Rafael Guzm\'an, Rob Kennicutt, David Koo, Trinh Thuan, Alexei Kniazev, Yuri Izotov,
Janice Lee, Jason Melbourne and Jose Herrero.  Finally, we wish to thank the  Astronomy 
Department of Case Western Reserve University for maintaining the Burrell Schmidt during 
the period of time when the survey observations reported here were obtained.


\appendix
\section{Supplementary Table of 4$\sigma$ Objects}

As we explained in Section 3, the main survey objects are selected based on the
presence of a 5$\sigma$ emission feature in their spectra. Because of the high 
sensitivity of the survey data, many objects were detected that have apparent 
emission lines with strengths that are only slightly weaker than the 5$\sigma$ limit. 
We decided to exclude such objects from the main survey list since one of the primary 
goals of the KISS project is to construct a deep but statistically complete sample of 
ELGs.  Early tests involving follow-up spectroscopy carried out on fields 
where objects were selected to lower thresholds showed that 5$\sigma$
detections were nearly always real sources, while objects between  4$\sigma$ 
and 5$\sigma$ tended to be real but also included a fair number ($\sim$25\%) of 
spurious sources.  However, these objects are nonetheless valid ELG candidates, and 
this list of objects likely includes a number of interesting objects.  Therefore, rather than 
ignore these weaker-lined ELG candidates entirely, we are publishing them in
a supplementary table.

Listed in Table \ref{table:tab3} are 158 ELG candidates that have emission lines 
detected at between the 4$\sigma$ and 5$\sigma$ level.  The format of Table 
\ref{table:tab3} is the same as for Table \ref{table:tab2}, except that the objects 
are now labeled with KISSRx numbers (`x' for extra).  The KISSRx numbers start at 
481 since we presented 480 KISSRx objects in KR1 and KR2. The full version of the table, 
as well as finder charts for all 158 KISSRx galaxies, are available in the electronic 
version of the paper. 

The supplementary ELG sample has characteristics similar to those of 
the main survey ELGs, although with some notable differences.  The median
H$\alpha$ equivalent width is 40.9 \AA, roughly 20\% lower than the value for 
the main sample.  The KISSRx galaxies are somewhat fainter (median B magnitude
of 18.65) and significantly redder (median B$-$V = 0.81).  Their median redshift
is slightly higher than that of either the Bo\"otes or Cetus field (0.067), and their median
luminosity is nearly a magnitude fainter ($-$18.23).  Hence, the supplementary ELG list
appears to be dominated by intermediate luminosity galaxies with somewhat
lower rates of star-formation activity (lower equivalent widths, redder colors)
than the ELGs in the main sample.   The differences between the KISSR and KISSRx
objects in the current paper are similar to those seen between the two samples
in KR1 and KR2.



\clearpage
\tablenum{1}
\begin{deluxetable}{lccc}
\tablenum{1}
\tablecaption{KISS NDWF Red Survey Observing Runs \label{table:tab1}}
\tablehead{
\colhead{Dates of Run} & 
\colhead{Number of} & 
\colhead{Number of} & 
\colhead{Number of} \\ 
\colhead{} &
\colhead{Nights\tablenotemark{a}} & 
\colhead{Fields -- Direct\tablenotemark{b}} & 
\colhead{Fields -- Spectral\tablenotemark{b}} \\
\colhead{(1)}&
\colhead{(2)}&
\colhead{(3)}&
\colhead{(4)}
}
\startdata
June 23 -- 24, 1998      & 2  & 8       & \nodata \\
September 17 -- 19, 1998 & 3  & \nodata & 10      \\
November 19 -- 23, 1998  & 3  & 12      & \nodata \\
May 12 -- 14, 1999       & 3  & \nodata & 8       \\
November 6, 1999         & 1  & \nodata & 2       \\
\enddata

\tablenotetext{a}{Number of nights during run that data were obtained.}

\tablenotetext{b}{Number of survey fields observed.}

\end{deluxetable}


\tablenum{2}
\begin{deluxetable}{rrrllcccrrcl}
\tablenum{2}
\tablecaption{List of Candidate ELGs\label{table:tab2}}
\tablewidth{0pt}
\tabletypesize{\scriptsize}
\tablehead{
\colhead{KISSR}&
\colhead{Field}&
\colhead{ID}&
\colhead{R.A.}&
\colhead{Dec.}&
\colhead{B}&
\colhead{B$-$V}&
\colhead{z$_{KISS}$}&
\colhead{Flux\tablenotemark{a}}&
\colhead{EW}&
\colhead{Qual.}&
\colhead{Comments}\\
\colhead{\#}&&&
\colhead{(J2000)}&
\colhead{(J2000)}&&&&&
\colhead{[\AA]}&& \\
\colhead{(1)}&
\colhead{(2)}&
\colhead{(3)}&
\colhead{(4)}&
\colhead{(5)}&
\colhead{(6)}&
\colhead{(7)}&
\colhead{(8)}&
\colhead{(9)}&
\colhead{(10)}&
\colhead{(11)}&
\colhead{(12)}
}
\startdata
 2158 & H0202 & 4755 & 2 01 59.3 & -4 41 03.5 & 17.88 & 0.59 & 0.0286 & 53 & 26 & 2 &\phm{imaveryveryverylongstring}\\
 2159 & G0202 & 3879 & 2 02 18.2 & -3 42 09.9 & 18.97 & 1.37 & 0.0871 & 34 & 21 & 3&\phm{imaveryveryverylongstring}\\
 2160 & G0202 & 3858 & 2 02 18.8 & -3 36 50.7 & 18.43 & 0.50 & 0.0415 & 72 & 64 & 1&\phm{imaveryveryverylongstring}\\
 2161 & G0202 & 3653 & 2 02 26.0 & -4 11 05.9 & 15.25 & 0.47 & 0.0159 & 353 & 50 & 1&\phm{imaveryveryverylongstring}\\
 2162 & G0202 & 3213 & 2 02 39.4 & -4 09 01.7 & 15.91 & 0.59 & 0.0465 & 316 & 28 & 2&\phm{imaveryveryverylongstring}\\
 2163 & G0202 & 3037 & 2 02 44.0 & -3 52 35.2 & 18.49 & 0.38 & 0.0143 &  46 & 36 & 2&\phm{imaveryveryverylongstring}\\
 2164 & G0202 & 3055 & 2 02 44.1 & -4 11 17.8 & 19.05 & 0.51 & 0.0430 &  39 & 60 & 2&\phm{imaveryveryverylongstring}\\
 2165 & G0202 & 2729 & 2 02 53.9 & -4 07 19.7 & 16.45 & 0.84 & 0.0435 & 138 & 14 & 1&\phm{imaveryveryverylongstring}\\
 2166 & G0202 & 2315 & 2 03 05.8 & -3 50 24.8 & 16.57 & 0.90 & 0.0427 &  38 & 4 & 3&\phm{imaveryveryverylongstring}\\
 2167 & G0202 & 2317 & 2 03 06.4 & -4 15 05.9 & 19.48 & 0.92 & 0.0811 &  39 & 47 & 2&\phm{imaveryveryverylongstring}\\ 
\\
 2168 & G0202 & 2329 & 2 03 06.5 & -4 27 14.1 & 17.11 & 0.41 & 0.0155 & 245 & 72 & 1&\phm{imaveryveryverylongstring}\\
 2169 & G0202 & 2135 & 2 03 13.2 & -4 18 47.3 & 19.01 & 0.83 & 0.0909 &  50 & 94 & 3&\phm{imaveryveryverylongstring}\\
 2170 & G0202 & 1836 & 2 03 21.4 & -4 03 31.0 & 18.05 & 0.87 & 0.0650 &  81 & 33 & 1&\phm{imaveryveryverylongstring}\\
 2171 & H0202 & 2122 & 2 03 25.5 & -5 04 24.4 & 20.01 & 0.51 & 0.0547 &  69 & 3054 & 3&\phm{imaveryveryverylongstring}\\
 2172 & G0202 & 1224 & 2 03 42.3 & -4 45 18.3 & 17.19 & 0.81 & 0.0370 &  54 & 9  & 3&\phm{imaveryveryverylongstring}\\
 2173 & G0202 & 1148 & 2 03 42.9 & -3 49 15.4 & 16.78 & 0.71 & 0.0469 & 410 & 57 & 1&\phm{imaveryveryverylongstring}\\
 2174 & G0202 & 786 & 2 03 54.0 & -3 53 00.1 & 15.76 & 0.63 & 0.0425 & 863 & 65  & 1&\phm{imaveryveryverylongstring}\\
 2175 & G0202 & 646 & 2 03 58.9 & -4 01 15.2 & 17.96 & 0.96 & 0.0617 &  68 & 24  & 1&\phm{imaveryveryverylongstring}\\
 2176 & G0202 & 550 & 2 04 00.9 & -3 16 08.8 & 21.00 & 1.68 & 0.0902 &  33 & 342 & 3&\phm{imaveryveryverylongstring}\\
 2177 & H0202 & 807 & 2 04 08.2 & -4 50 47.1 & 17.78 & 0.53 & 0.0771 & 117 & 97  & 1&\phm{imaveryveryverylongstring}\\ 
\\
 2178 & G0202 & 420 & 2 04 08.5 & -4 49 15.4 & 18.13 & 1.17 & 0.0829 &  45 & 25 & 2&\phm{imaveryveryverylongstring}\\
 2179 & G0202 & 372 & 2 04 09.5 & -4 28 20.2 & 18.33 & 0.79 & 0.0793 &  89 & 54 & 1&\phm{imaveryveryverylongstring}\\
 2180 & H0202 & 549 & 2 04 16.9 & -4 48 41.0 & 17.29 & 0.70 & 0.0869 &  77 & 24 & 1&\phm{imaveryveryverylongstring}\\
 2181 & G0202 & 76  & 2 04 17.0 & -3 39 15.2 & 19.17 & 0.90 & 0.0777 &  66 & 101 & 2&\phm{imaveryveryverylongstring}\\
 2182 & H0202 & 434 & 2 04 20.3 & -4 54 58.1 & 14.99 & 0.28 & 0.0178 & 151 & 32 & 1&\phm{imaveryveryverylongstring}\\
 2183 & G0205 & 5741 & 2 04 21.7 & -3 42 40.0 & 17.71 & 0.83 & 0.0586 &  54 & 22 & 1&\phm{imaveryveryverylongstring}\\
 2184 & G0205 & 5425 & 2 04 30.4 & -3 36 09.1 & 17.71 & 0.76 & 0.0558 &  72 & 23 & 1&\phm{imaveryveryverylongstring}\\
 2185 & G0205 & 5416 & 2 04 31.4 & -4 03 34.6 & 19.56 & 0.67 & 0.0387 &  59 & 168 & 1&\phm{imaveryveryverylongstring}\\
 2186 & G0205 & 5375 & 2 04 33.4 & -4 29 39.6 & 17.83 & 0.45 & 0.0483 &  97 & 65 & 1&\phm{imaveryveryverylongstring}\\
 2187 & H0205 & 6171 & 2 04 35.9 & -5 30 00.0 & 18.12 & 0.78 & 0.0764 & 151 & 92 & 1&\phm{imaveryveryverylongstring}\\

\enddata
\tablenotetext{\,}{Note. -- The complete version of this table is presented in 
the electronic edition of the Journal.  A portion is shown here for guidance 
regarding its content and format.}
\tablenotetext{a}{Units of 10$^{-16}$ erg/s/cm$^2$}
\end{deluxetable}


\tablenum{3}
\begin{deluxetable}{ccccccccc}
\tablenum{3}
\tablecaption{$V/V_{max}$ Test \label{table:comptab}}
\tablewidth{0pt}
\tabletypesize{\scriptsize}
\tablehead{
\colhead{} &
\colhead{Total}  &
\colhead{Number} &
\colhead{Number} &
\colhead{} &
\colhead{Number} &
\colhead{Cumulative} &
\colhead{\%} &
\colhead{\%} \\
\colhead{$m_L$} &
\colhead{Number} &
\colhead{Flux} &
\colhead{Volume} &
\colhead{$\langle V/V_{max}\rangle$} &
\colhead {added} &
\colhead{number} &
\colhead{Flux} &
\colhead{Complete}\\
\colhead{} &
\colhead{} &
\colhead{Limited} &
\colhead{Limited} &
\colhead{} &
\colhead{} &
\colhead{added} &
\colhead{Limited} &
\colhead{} \\
\colhead{(1)} &
\colhead{(2)} &
\colhead{(3)} &
\colhead{(4)} &
\colhead{(5)} &
\colhead{(6)} &
\colhead{(7)} &
\colhead{(8)} &
\colhead{(9)}
}
\startdata
13.0 & \phn\phn 4 & \phn\phn 4 & \phn\phn 0 & 0.5922 & \phn 0 & \phn\phn 0 & 100.00 & 100.00 \\ 
13.1 & \phn\phn 5 & \phn\phn 5 & \phn\phn 0 & 0.5900 & \phn 0 & \phn\phn 0 & 100.00 & 100.00 \\
13.2 & \phn\phn 5 & \phn\phn 5 & \phn\phn 0 & 0.5139 & \phn 0 & \phn\phn 0 & 100.00 & 100.00 \\
13.3 & \phn\phn 7 & \phn\phn 7 & \phn\phn 0 & 0.5889 & \phn 0 & \phn\phn 0 & 100.00 & 100.00 \\
13.4 & \phn\phn 8 & \phn\phn 8 & \phn\phn 0 & 0.5688 & \phn 0 & \phn\phn 0 & 100.00 & 100.00 \\
13.5 & \phn\phn 9 & \phn\phn 9 & \phn\phn 0 & 0.5468 & \phn 0 & \phn\phn 0 & 100.00 & 100.00 \\
13.6 & \phn 12 & \phn 12 & \phn\phn 0 & 0.5928 & \phn 0 & \phn\phn 0 & 100.00 & 100.00 \\  
13.7 & \phn 16 & \phn 16 & \phn\phn 0 & 0.6271 & \phn 0 & \phn\phn 0 & 100.00 & 100.00 \\ 
13.8 & \phn 20 & \phn 20 & \phn\phn 0 & 0.6256 & \phn 0 & \phn\phn 0 & 100.00 & 100.00 \\ 
13.9 & \phn 22 & \phn 20 & \phn\phn 2 & 0.5666 & \phn 0 & \phn\phn 0 & \phn 90.91 & 100.00 \\ 
14.0 & \phn 28 & \phn 24 & \phn\phn 4 & 0.6046 & \phn 0 & \phn\phn 0 &  \phn 85.71 & 100.00 \\ 
14.1 & \phn 35 & \phn 28 & \phn\phn 7 & 0.6352 & \phn 0 & \phn\phn 0 & \phn 80.00 & 100.00 \\ 
14.2 & \phn 38 & \phn 31 & \phn\phn 7 & 0.5891 & \phn 0 & \phn\phn 0 & \phn 81.58 & 100.00 \\ 
14.3 & \phn 44 & \phn 36 & \phn\phn 8 & 0.5833 & \phn 0 & \phn\phn 0 & \phn 81.82 & 100.00 \\ 
14.4 & \phn 54 & \phn 42 & \phn 12 & 0.5818 & \phn 0 & \phn\phn 0 & \phn 77.78 & 100.00 \\ 
14.5 & \phn 67 & \phn 50 & \phn 17 & 0.5946 & \phn 0 & \phn\phn 0 & \phn 74.63 & 100.00 \\ 
14.6 & \phn 81 & \phn 56 & \phn 25 & 0.6012 & \phn 0 & \phn\phn 0 & \phn 69.14 & 100.00 \\ 
14.7 & \phn 98 & \phn 67 & \phn 31 & 0.6173 & \phn 0 & \phn\phn 0 & \phn 68.37 & 100.00 \\ 
14.8 &  108 & \phn 69 & \phn 39 & 0.5764 & \phn 0 & \phn\phn 0 & \phn 63.89 & 100.00 \\ 
14.9 &  120 & \phn 76 & \phn 44 & 0.5646 & \phn 0 & \phn\phn 0 & \phn 63.33 & 100.00 \\ 
15.0 &  139 & \phn 87 & \phn 52 & 0.5642 & \phn 0 & \phn\phn 0 & \phn 62.59 & 100.00 \\ 
15.1 &  154 & \phn 93 & \phn 61 & 0.5480 & \phn 0 & \phn\phn 0 & \phn 60.39 & 100.00 \\ 
15.2 &  166 & \phn 99 & \phn 67 & 0.5251 & \phn 0 & \phn\phn 0 & \phn 59.64 & 100.00 \\ 
15.3 &  184 & 106 & \phn 78 & 0.5122 & \phn 0 & \phn\phn 0 & \phn 57.61 & 100.00 \\ 
15.4 &  195 & 107 & \phn 88 & 0.4822 & \phn 4 & \phn\phn 4 & \phn 54.87 & \phn 96.40 \\ 
15.5 &  206 & 114 & \phn 92 & 0.4612 & \phn 6 & \phn 10 & \phn 55.34 & \phn 91.94 \\ 
15.6 &  216 & 117 & \phn 99 & 0.4330 & \phn 9 & \phn 19 & \phn 54.17 & \phn 86.03 \\ 
15.7 &  225 & 120 & 105 & 0.4104 & 10 & \phn 29 & \phn 53.33 & \phn 80.54 \\ 
15.8 &  231 & 119 & 112 & 0.3821 & 13 & \phn 42 & \phn 51.52 & \phn 73.91 \\ 
15.9 &  238 & 117 & 121 & 0.3761 & 10 & \phn 52 & \phn 49.16 & \phn 69.23 \\ 
16.0 &  244 & 110 & 134 & 0.3417 & 17 & \phn 69 & \phn 45.08 & \phn 61.45 \\ 
16.1 &  253 & 104 & 149 & 0.3333 & 13 & \phn 82 & \phn 41.11 & \phn 55.91 \\ 
16.2 &  257 & \phn 94 & 163 & 0.3116 & 16 & \phn 98 & \phn 36.58 & \phn 48.96 \\ 
16.3 & 259 & \phn 88 & 171 & 0.2757 & 21 & 119 & \phn 33.98 & \phn 42.51 \\ 
16.4 & 260 & \phn 83 & 177 & 0.2521 & 22 & 141 & \phn 31.92 & \phn 37.05 \\ 
16.5 & 261 & \phn 76 & 185 & 0.2258 & 24 & 165 & \phn 29.12 & \phn 31.54 \\ 
\enddata

\end{deluxetable}


\tablenum{4}
\begin{deluxetable}{rrrllcccrrcl}
\tablecaption{List of 4$\sigma$ Candidate ELGs\label{table:tab3}}
\tablewidth{0pt}
\tabletypesize{\scriptsize}
\tablehead{
\colhead{KISSRx}&
\colhead{Field}&
\colhead{ID}&
\colhead{R.A.}& 
\colhead{Dec.}&
\colhead{B}&
\colhead{B$-$V}&
\colhead{z$_{KISS}$}& 
\colhead{Flux\tablenotemark{a}}&
\colhead{EW}&
\colhead{Qual.}& 
\colhead{Comments} \\
\colhead{\#}&&&
\colhead{(J2000)}&
\colhead{(J2000)}&&&&&
\colhead{[\AA]}&& \\
\colhead{(1)}&
\colhead{(2)}&
\colhead{(3)}&
\colhead{(4)}&
\colhead{(5)}&
\colhead{(6)}&
\colhead{(7)}&
\colhead{(8)}&
\colhead{(9)}&
\colhead{(10)}&
\colhead{(11)}&
\colhead{(12)}
}
\startdata
481 & G0202 & 5949 & 2 01 17.5 & -3 54 59.0 & 18.12 & 0.57 & 0.0396 & 43 & 19 & 2 &\phm{imaveryveryverylongstring}\\
482 & G0202 & 5213 & 2 01 41.7 & -4 16 42.6 & 21.07 & 1.88 & 0.0816 & 79 & 116 & 3 &\phm{imaveryveryverylongstring}\\
483 & G0202 & 4890 & 2 01 49.8 & -3 46 50.7 & 18.98 & 0.77 & 0.0729 & 92 & 71 & 2 &\phm{imaveryveryverylongstring}\\
484 & G0202 & 4563 & 2 01 57.6 & -3 17 19.8 & 19.27 & 0.76 & 0.0407 & 29 & 43 & 3 &\phm{imaveryveryverylongstring}\\
485 & G0202 & 4294 & 2 02 07.1 & -4 05 04.2 & 19.13 & 1.05 & 0.0597 & 59 & 32 & 2 &\phm{imaveryveryverylongstring}\\
486 & G0202 & 4158 & 2 02 11.2 & -4 07 33.9 & 17.76 & 1.03 & 0.0849 & 50 & 19 & 3 &\phm{imaveryveryverylongstring}\\
487 & G0202 & 2393 & 2 03 04.4 & -4 28 15.8 & 17.89 & 1.03 & 0.0370 & 51 & 18 & 2 &\phm{imaveryveryverylongstring}\\
488 & G0202 & 2057 & 2 03 13.9 & -3 19 42.6 & 20.58 & 0.57 & 0.0767 & 49 & 228 & 3 &\phm{imaveryveryverylongstring}\\
489 & G0202 & 1608 & 2 03 27.2 & -3 17 14.9 & 17.74 & 0.15 & 0.0375 & 46 & 32 & 2 &\phm{imaveryveryverylongstring}\\
490 & G0202 & 1499 & 2 03 31.6 & -3 46 56.4 & 19.07 & 0.70 & 0.0873 & 39 & 33 & 3 &\phm{imaveryveryverylongstring}\\
\\
491 & H0202 & 1892 & 2 03 32.0 & -4 21 50.5 & 20.42 & 1.16 & 0.0515 & 64 & 105 & 2 &\phm{imaveryveryverylongstring}\\
492 & H0202 & 1443 & 2 03 46.9 & -4 35 15.3 & 19.10 & 0.82 & 0.0546 & 43 & 6 & 3 &\phm{imaveryveryverylongstring}\\
493 & H0202 & 749 & 2 04 09.5 & -4 28 19.8 & 18.34 & 0.75 & 0.0733 & 121 & 69 & 2 &\phm{imaveryveryverylongstring}\\
494 & G0205 & 4694 & 2 04 53.7 & -4 00 57.6 & 19.44 & 0.24 & 0.0135 & 31 & 111 & 3 &\phm{imaveryveryverylongstring}\\
495 & G0205 & 4052 & 2 05 12.1 & -3 28 55.0 & 19.35 & 0.74 & 0.0674 & 93 & 50 & 2 &\phm{imaveryveryverylongstring}\\
496 & G0205 & 3412 & 2 05 33.2 & -3 57 25.3 & 18.78 & 0.65 & 0.0926 & 44 & 54 & 2 &\phm{imaveryveryverylongstring}\\
497 & G0205 & 3002 & 2 05 45.8 & -3 15 37.4 & 23.42 & 3.92 & 0.0713 & 67 & 84 & 3 &\phm{imaveryveryverylongstring}\\
498 & G0205 & 2790 & 2 05 52.7 & -3 23 46.0 & 18.30 & 0.82 & 0.0827 & 20 & 11 & 3 &\phm{imaveryveryverylongstring}\\
499 & H0205 & 3297 & 2 06 00.2 & -4 40 31.8 & 23.69 & 3.69 & 0.0673 & 40 & 322 & 3 &\phm{imaveryveryverylongstring}\\
500 & H0205 & 3071 & 2 06 08.5 & -5 20 22.8 & 19.37 & 0.73 & 0.0744 & 32 & 35 & 2 &\phm{imaveryveryverylongstring}\\
\\
501 & G0205 & 2173 & 2 06 14.5 & -4 28 14.9 & 18.16 & 0.13 & 0.0341 & 53 & 45 & 2 &\phm{imaveryveryverylongstring}\\
502 & G0205 & 1969 & 2 06 21.7 & -4 13 34.2 & 16.73 & 0.67 & 0.0346 & 75 & 11 & 2 &\phm{imaveryveryverylongstring}\\
503 & G0205 & 1809 & 2 06 25.4 & -3 29 24.8 & 21.11 & 1.16 & 0.0684 & 28 & 57 & 2 &\phm{imaveryveryverylongstring}\\
504 & G0205 & 1512 & 2 06 35.9 & -3 38 17.3 & 16.35 & 0.27 & 0.0284 & 115 & 12 & 3 &\phm{imaveryveryverylongstring}\\
505 & G0205 & 1496 & 2 06 37.6 & -4 19 43.9 & 19.82 & 1.48 & 0.0388 & 30 & 19 & 3 &\phm{imaveryveryverylongstring}\\
506 & H0205 & 1656 & 2 06 50.8 & -5 32 50.7 & 18.52 & 0.84 & 0.0245 & 32 & 19 & 3 &\phm{imaveryveryverylongstring}\\
507 & G0205 & 943 & 2 06 54.0 & -3 35 21.2 & 19.68 & 1.20 & 0.0682 & 40 & 28 & 2 &\phm{imaveryveryverylongstring}\\
508 & H0205 & 1333 & 2 07 01.1 & -5 47 59.0 & 18.66 & 0.66 & 0.0345 & 44 & 32 & 3 &\phm{imaveryveryverylongstring}\\
509 & G0205 & 417 & 2 07 10.9 & -3 30 36.5 & 19.50 & 0.89 & 0.0763 & 84 & 362 & 2 &\phm{imaveryveryverylongstring}\\
510 & G0205 & 52 & 2 07 21.7 & -3 15 56.5 & 19.89 & 0.81 & 0.0091 & 52 & 64 & 3 &\phm{imaveryveryverylongstring}\\
\enddata

\tablenotetext{\,}{Note.--- The complete version of this table is presented in the
electronic edition of the Journal.  A portion is shown here for guidance regarding
its content and format.}

\tablenotetext{a}{Units of 10$^{-16}$ erg/s/cm$^2$}

\end{deluxetable}


\end{document}